\documentclass{article}
\usepackage{graphicx}
\begin{document}
\title{The Image of a Wormhole}
\author{Frank Y. Wang \\
Department of Mathematics \\
LaGuardia Community College of the City
University of New York \\
30-10 Thomson Avenue Long Island City New York 11101 \\
\texttt{fwang@lagcc.cuny.edu}}
\date{}
\maketitle

\section*{Introduction}

A wormhole is a hypothetical tunnel through space. Such a concept has
long been a favorite theme in \emph{Star Trek} and in science fiction.
An extensive list of popular science books\cite{science} are devoted
to topics related to wormholes.  We might gauge the popularity of
wormholes among students by the recurring references to them in
\emph{The Simpsons}: in one Halloween special\cite{simp}, Homer was
sucked into modern-day California while saying ``there's so much I
don't know about astrophysics, I wish I'd read that book by
[Dr. Hawking].''  The detailed calculations about a wormhole are
complex and still under debate among physicists, but with the
techniques taught in a standard calculus course\cite{calculus}, a
student is capable of producing the images of wormholes based on sound
physical principles.  Because many issues concerning wormholes are not
fully resolved yet, this paper is not meant to be rigorous.  We merely
attempt to motivate students to apply mathematical skills that they
have learned in calculus, and to induce their interest in
non-Euclidean geometry and general relativity through this fascinating
subject.

\section{Einstein's Theory of Gravity}

Newton's law of gravity states that the gravitational force between
two bodies is $F = G M m/r^{2}$, where $G$ is the gravitational
constant, $m$ and $M$ are the masses of the two bodies, and $r$ is the
distance between them.  Einstein had an entirely different view of
gravity.  Suppose we were to measure the length of the equator of the
earth very carefully and obtain the circumference $C$, we would
predict that the radius of the earth is $C/2 \pi$ based on Euclidean
geometry.  However, if we indeed dig a hole to the center of the earth
and measure the radius, we would have a value greater (about 1.5
millimeters) than our prediction.  If we do the same for the sun, the
discrepancy would be about one-half a kilometer\cite{feyn}.

\begin{figure}[h]
\includegraphics[scale=0.8]{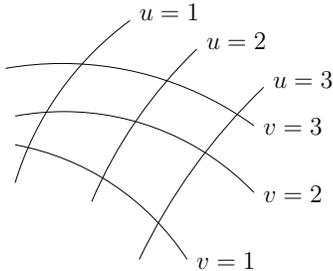}
\caption{Gaussian coordinates.}
\label{fig:gauss}
\end{figure}

From this thought experiment, we conclude that the space is curved.
Therefore, the differential of arc length based on Euclidean geometry,
\begin{equation}
d \sigma^{2} = d x^{2} + d y^{2}  ,
\end{equation}
is no longer valid.  Gauss has developed a generalization of the
Cartesian coordinates by drawing a system of arbitrary curves on the
surface, designated as $u=1$, $u=2$, $u=3$, $\ldots$, $v=1$, $v=2$,
$v=3$, and so forth (see Figure~\ref{fig:gauss}).

Using the Gaussian coordinates, the differential of arc length is
\begin{equation}
d \sigma^{2} = g_{11} \, d u^{2} + 2 g_{12} \, d u \, d v + 
g_{22} \, d v^{2} ,
\end{equation}
where the coefficients $g_{\mu \nu}$, called metric coefficients, are
constants or functions of $u$ and $v$.  This equation can be construed
as a generalized ``Pythagorean theorem.''  As an example, with the polar
coordinates $(r, \phi)$, we have
\begin{equation}
d \sigma^{2} = d r^{2} + r^{2} d \phi^{2} ,
\end{equation}
where the metric coefficients are $g_{rr}=1$, $g_{\phi \phi}=r^{2}$
and $g_{r \phi}=0$ (because the system is orthogonal).  Riemann
extended the Gaussian geometry to greater number of dimensions, and
Riemannian geometry later became the mathematical framework for
Einstein's general relativity.  In brief, Einstein conceived of
gravity as a change in the geometry of spacetime due to mass-energy,
and his greatest achievement was to derive a field equation which
relates the metric coefficients $g_{\mu \nu}$ to
mass-energy\cite{mtw}.

\section{Schwarzschild Wormhole and Embedding Diagram}

Shortly after Einstein published his general theory of relativity,
Schwarzschild found a solution\cite{sch} for the exterior of a
spherically symmetric gravitational source $M$.  (A solution to
Einstein's field equation can be written as a differential of arc
length which describes the spacetime warped by gravity.) In the
equatorial plane at a fixed time, the solution is simplified to
\begin{equation}\label{eq:sch}
d \sigma^{2} = \left(1 - \frac{2 M}{r} \right)^{-1} d r^{2}
+ r^{2} d \phi^{2} .
\end{equation}
We have adopted a convention in general relativity of measuring mass
in units of length by multiplying the ordinary mass in kilogram by a
factor of $G/c^{2}$, where $G$ is the gravitational constant and $c$
the speed of light.  For example, the mass of the earth is 0.444 cm,
and that of the sun 1.48 km.  A reader may notice the presence of a
singularity at $r=2M$ in equation~(\ref{eq:sch}).  If we compress the
sun into a sphere of a radius less than $2 \times 1.48$ km, we will
have a black hole.  The physical significance of this singularity is
that nothing, not even light, can escape the gravitational pull of the
black hole once it crosses the boundary $r=2M$ (the event horizon).

\begin{figure}[h]
\includegraphics[scale=0.6]{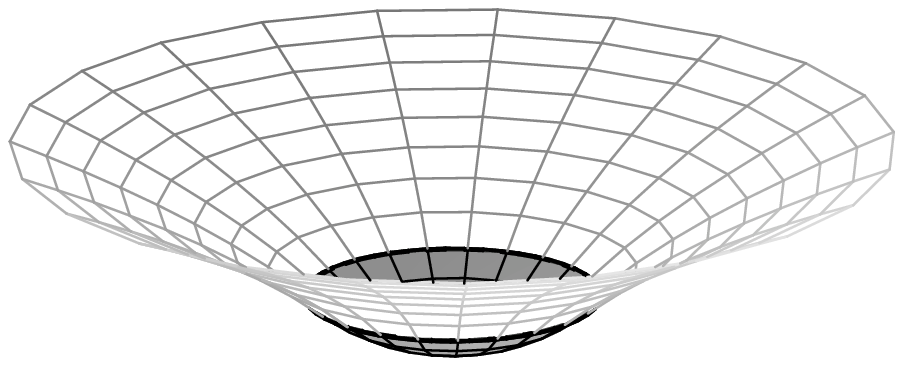}
\hspace{0.1in}
\includegraphics[scale=0.6]{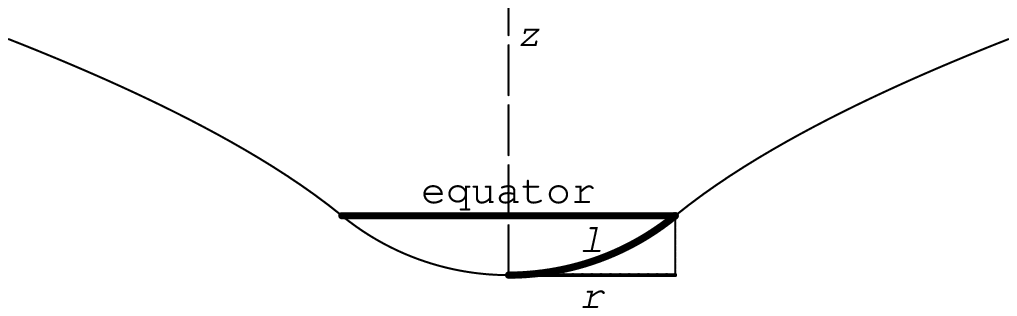}
\caption{An embedding diagram of the equatorial plane of a massive
  spherical object.  The measured radius is $l$, while the length of
  equator divided by $2 \pi$ gives $r$.}
\label{fig:fig1}
\end{figure}

Equation (\ref{eq:sch}) is a two-dimensional curved surface, and it
can be visualized by means of an embedding diagram\cite{emb}.  As
mentioned earlier, the measured radius in a curved space is greater
than the ratio $C/2 \pi$ (the supposed radius in a flat space).  To
visualize this phenomenon, we imagine a fictitious depth $z$, see
Figure~\ref{fig:fig1}, to accommodate the actual radius.  We emphasize
that this artificial $z$ dimension is purely for visual purpose and
has nothing to do with real space.

To be more quantitative, the measured distance in a curved space is
the integral $\int d \sigma$.  We want to find an embedding formula
$z$, such that the geometry of a curved two-dimensional space is the
same as a flat three-dimensional space.  A flat (Euclidean)
three-dimensional space means that the ordinary Pythagorean theorem is
valid, or
$$
d \sigma^{2} = d x^{2} + d y^{2} + d z^{2} .
$$ 
It is more convenient to rewrite this differential of arc length
using cylindrical coordinates:
\begin{equation}\label{eq:cyl}
d \sigma^{2} = d z^{2} + d r^{2} + r^{2} d \phi^{2} .
\end{equation}
Comparing equations (\ref{eq:sch}) and (\ref{eq:cyl}), we can solve
for $dz$:
\begin{equation}
d z = 
\pm \sqrt{ \left(1 - \frac{2 M}{r} \right)^{-1} - 1 } 
\ dr
= \sqrt{\frac{2 M}{r - 2 M}} \, dr ,
\end{equation}
and integrate to obtain $z$:
\begin{equation}
z = \pm \int \sqrt{\frac{2 M}{r - 2 M}} \, d r 
= \pm 2 \sqrt{2 M} \sqrt{r - 2 M} . 
\end{equation}
Because of spherical symmetry, there is no explicit $\phi$ term.  To
restore the $\phi$ dimension, we rotation the curve $z$ around the
$z$-axis, which gives the embedding surface.  In parametric form, the
embedding surface is written as $(r \cos \phi, r \sin \phi,
\sqrt{8M(r-2M)})$; computer programs are readily available for
graphing this surface\cite{maple}.

\begin{figure}[h]
\includegraphics[scale=0.5]{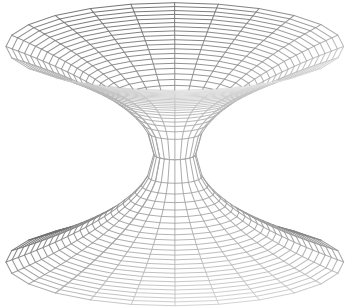}
\caption{The Einstein-Rosen bridge, or the Schwarzschild wormhole,
  embedded in a three-dimensional Euclidean space.}
\label{fig:fig2}
\end{figure}

Einstein and Rosen\cite{ein} proposed a topology that connects two
universes at $r=2M$, and they called such a connection a ``bridge.''
By joining the positive and negative solution of $z$ at $r=2M$, we
have the embedding diagram for the Einstein-Rosen bridge, as shown in
Figure~\ref{fig:fig2}.  Based of its shape, J. A. Wheeler coined the
term ``wormhole'' for this type of geometry.

\section{Morris-Thorne Wormhole}

In 1985, Carl Sagan wrote a novel \emph{Contact}\cite{sagan}, which
was later adopted to a film of the same title released in 1997.
Before Sagan published the book, he sought advice about gravitational
physics from a Caltech physicist, Kip Thorne.  In the original
manuscript, Sagan had his heroine, Eleanor Arroway (played by Jodie
Foster in the film), plunge into a black hole near the earth, travel
through the space, and emerge an hour later near the star Vega, 26
light-years away.  Thorne pointed out a well-established result that
it is impossible to travel through the Schwarzschild wormhole because
its throat pinches off too quickly\cite{thorne}; in other words,
Figure~\ref{fig:fig2} exists only for a brief moment which is too
short to allow communicating with or traveling to the other part of
the universe.

Sagan's request, however, piqued Thorne's curiosity about wormholes.
Thorne devised a wormhole solution\cite{mt88}, which is simplified in
the equatorial plane at a fixed time as
\begin{equation}\label{eq:mt}
d \sigma^{2} = 
\frac{1}{1-b_{0}^{2}/r^{2}} d r^{2} + r^{2} d \phi^{2} ,  
\end{equation}
where $b_{0}$ is the radius of the throat.  

It is easy to write a mathematical solution for a geometry that we
desire, but Einstein's field equation relates geometry to mass-energy.
If we attempt to construct a wormhole which has a geometry as
equation~(\ref{eq:mt}) and remains open and stable so that it allows
two-way travel, we will need negative-energy material, called
\emph{exotic matter} by Thorne (and incorporated into Sagan's novel).
There is a debate about the possibility of such exotic matter, but it
is certain that the energy required is far beyond the producing
capacity of a present and foreseeable future civilization.  The hope
of space travel in short term is impractical at least, if not entirely
impossible.

\begin{figure}[h]
\includegraphics[scale=0.5]{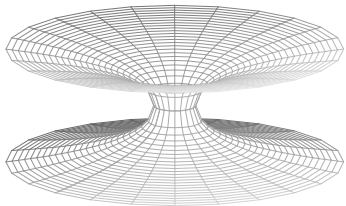}
\caption{The Morris-Thorne wormhole embedded in a three-dimensional
  Euclidean space.}
\label{fig:fig3}
\end{figure}

Nevertheless, we employ the same procedure as the preceding section to
derive the embedding formula to visualize this geometry.  From
equations (\ref{eq:mt}) and (\ref{eq:cyl}), we solve for $dz$:
\begin{equation}
dz = \pm \sqrt{\frac{1}{1 - b_{0}^{2}/r^{2}} - 1} \ dr
= \pm \sqrt{\frac{b_{0}^{2}}{r^{2} - b_{0}^{2}}} \, dr.
\end{equation}  
With a substitution $r=b_{0} \sec \theta$, we integrate $dz$ to obtain
\begin{equation}
z = \pm b_{0} 
\ln \left[\frac{r}{b_{0}} + 
\sqrt{\left( \frac{r}{b_{0}} \right)^{2} -1} \right]
\end{equation}
Alternatively, with a substitution $r=b_{0} \cosh \theta$, we obtain
\begin{equation}
z = \pm b_{0} \cosh^{-1} \left( \frac{r}{b_{0}} \right) ,
\end{equation}
which is a catenary curve.  The surface of revolution of this curve is
shown in Figure~\ref{fig:fig3}.

\section{Summary}
The images that appear in science literature depicting how a massive
body warps space are based on the concept of embedding a curved
two-dimensional surface in a three-dimensional flat (Euclidean) space.
Quantitatively, the Schwarzschild wormhole can be visualized as a
paraboloid of revolution
\begin{equation}
r = 2 M + \frac{z^{2}}{8 M} ,
\end{equation}
and the Morris-Thorne wormhole as a catenoid of revolution
\begin{equation}
r = b_{0} \cosh \left(\frac{z}{b_{0}} \right) .
\end{equation}

\section*{Acknowledgments}
I thank Dr. Gordon Crandall and Dr. Kamal Hajallie for useful
suggestions.

\end{document}